\documentclass[
amsfonts,
amsmath,
aps,
prd,
nofootinbib,
reprint
]{revtex4-2}
\usepackage[dvipdfmx]{graphicx}
\usepackage{bm,latexsym,amsmath,amssymb,amsfonts} 
\DeclareMathAlphabet{\mathpzc}{OT1}{pzc}{m}{it}
\usepackage[normalem]{ulem}
\usepackage[breaklinks=true,colorlinks=true]{hyperref}
\usepackage{xcolor}
\usepackage{wasysym}
\usepackage[mathscr]{eucal}
\usepackage{multirow}
\usepackage{enumitem}
\usepackage{tabularray}
\usepackage{tablefootnote}
\usepackage{longtable}
\usepackage{booktabs}
\usepackage{lipsum}

\newcommand{\orcid}[1]{\href{https://orcid.org/#1}{\includegraphics[width=8pt]{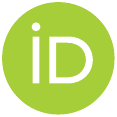}}}
\definecolor{royalblue}{rgb}{0.0, 0.0, 0.8}
\hypersetup{linkcolor=royalblue, citecolor=royalblue, urlcolor=royalblue}

\begin{document}

\def\Journal#1#2#3{\href{https://doi.org/#3}{{#1} #2}}
\def\arXiv#1#2{\href{https://arxiv.org/abs/#1}{arXiv:#1}}
\def\supplmat{\hyperref[sec. supplementary material]{Supplementary Material}}
        
\title{Two descriptions of dark matter around a black hole:\\ photon sphere, shadow, and lensing}

\author{M.~F.~Fauzi\orcid{0009-0005-3380-6005}}
\thanks{Corresponding author.}
\email{muhammad.fahmi31@ui.ac.id}
\affiliation{Departemen Fisika, FMIPA, Universitas Indonesia, Depok 16424, Indonesia}

\author{H.~S.~Ramadhan\orcid{0000-0003-0727-738X}}
\email{hramad@sci.ui.ac.id}
\affiliation{Departemen Fisika, FMIPA, Universitas Indonesia, Depok 16424, Indonesia}

\author{A.~Sulaksono\orcid{0000-0002-1493-5013}}
\email{anto.sulaksono@sci.ui.ac.id}
\affiliation{Departemen Fisika, FMIPA, Universitas Indonesia, Depok 16424, Indonesia}


\begin{abstract}
We examine the observational discrepancies of two widely used models describing anisotropic (dark) matter distributions around a black hole, focusing on their photon spheres, shadow radii, and lensing observables. The models considered are the vacuum and Einstein cluster dark matter models, characterized by negative and zero radial pressure, respectively. The analysis reveals that these models display contrasting photon sphere behaviors. In particular, the Einstein cluster results in a more pronounced deviation in the shadow radius relative to the standard Schwarzschild black hole. Additionally, a distinctive lensing phenomenon associated with the matter halo is identified in both models.
\end{abstract}

\maketitle
\section{Introduction}
Black holes (BHs) at galactic centers are expected to be surrounded by dark matter (DM) distributions~\cite{Navarro:1996gj,Gondolo:1999ef,Sadeghian:2013laa,Bertone:2004pz}. A wide range of astrophysical observations provides strong evidence for the presence of DM in galaxies (see, e.g.,~\cite{Clowe:2006eq,Bertone:2016nfn}). Motivated by this, numerous theoretical studies have explored the influence of DM-surrounded BHs, including the observational signatures from such systems~\cite{Toshmatov:2025rln,Jusufi:2019ltj,Macedo:2024qky,Ali:2025ney,Konoplya:2025ect,Konoplya:2025nqv,Konoplya:2022hbl,Xu:2018wow,Cardoso:2021wlq,Myung:2025cxu,Konoplya:2019sns,Xavier:2023exm,Hou:2018bar,Ma:2022jsy,Kouniatalis:2025itj,Speeney:2024mas,Figueiredo:2023gas,Gliorio:2025cbh,Liu:2021xfb,Jusufi:2020cpn,Liu:2024qso,Xu:2017bpz,He:2024amh,Pantig:2022toh,Liu:2023xtb,Molla:2025yoh,Molla:2023yxn,Jusufi:2022jxu}. These studies typically use phenomenological approaches to model DM distributions \textemdash \textit{i.e.}, using empirical DM models (e.g.~\cite{Hernquist:1990be,Dehnen:1993uh,Navarro:1996gj,Einasto:1996}) instead of solutions derived from fundamental principles\textemdash some of which were constructed based on N-body simulations of galaxies.

However, there is still no consensus on how DM modifies the spacetime geometry in the vicinity of a BH. A number of studies adopt the simplifying assumption of a vacuum spacetime~\cite{Jusufi:2019ltj,Toshmatov:2025rln,Xu:2018wow,Konoplya:2025ect,Hou:2018bar,Konoplya:2019sns,Jusufi:2020cpn,Pantig:2022toh,Liu:2023xtb,Molla:2025yoh,Molla:2023yxn,Liu:2021xfb,Xu:2017bpz,Liu:2024qso,He:2024amh,Myung:2025cxu,Ali:2025ney,Ma:2022jsy}, 
\begin{equation}
\label{gtt}
-g_{tt} = g_{rr}^{-1},
\end{equation}
effectively modeling the region around the BH as being oblivious to the presence of DM. It has been argued that this approximation captures the essential physical behavior of more general treatments~\cite{Xu:2018wow}. This approach was recently criticized in Ref.~\cite{Bolokhov:2025zva}, as the metric is inconsistent with the intended matter distribution. In contrast, Cardoso \textit{et al.}~\cite{Cardoso:2021wlq} used the \textit{Einstein cluster} to model the (dark) matter distributions around a BH in a galaxy, which leads to a spacetime geometry that explicitly violates condition \eqref{gtt} due to the zero radial pressure matter distribution. This distinction raises the question: Does the simplified metric approximation adequately capture the observational signatures expected from a more complete Einstein cluster description of DM in the strong field regime?

This letter addresses the above question from the perspective of electromagnetic observations. While other studies focus on a single DM description, our work compares the two models and identifies their key distinguishing features. We consider two observational regimes. First, we examine the BH shadow as a probe of the strong-field region. Although numerous studies have examined this context~\cite{Myung:2025cxu,Konoplya:2019sns,Hou:2018bar,Jusufi:2020cpn,Ma:2022jsy,Ali:2025ney,Jusufi:2022jxu,Xavier:2023exm,Macedo:2024qky,Konoplya:2025nqv}, most of them employ the assumption~\eqref{gtt}\textemdash or an equivalent condition leading to~\eqref{gtt}~\cite{Myung:2025cxu,Konoplya:2019sns,Hou:2018bar,Jusufi:2020cpn,Ma:2022jsy,Ali:2025ney}. This approach may provide insight into the characteristics of galactic DM inferred from the shadow of the central BH, particularly given observational results from the Event Horizon Telescope~\cite{EventHorizonTelescope:2019dse,EventHorizonTelescope:2019uob,EventHorizonTelescope:2019jan,EventHorizonTelescope:2019ths,EventHorizonTelescope:2019pgp,EventHorizonTelescope:2019ggy,EventHorizonTelescope:2021bee,EventHorizonTelescope:2021srq,EventHorizonTelescope:2023gtd,EventHorizonTelescope:2022wkp,EventHorizonTelescope:2022apq,EventHorizonTelescope:2022wok,EventHorizonTelescope:2022exc,EventHorizonTelescope:2022urf,EventHorizonTelescope:2022xqj,EventHorizonTelescope:2024hpu,EventHorizonTelescope:2024rju}. Second, motivated by the fact that DM distributions typically span over galactic scales, we investigate a complementary observable in a weaker-field regime\textemdash namely, the image positions and time delays arising from gravitational lensing, following the formalism of Virbhadra~\cite{Virbhadra:2008ws}. This choice was motivated by the findings of Boehmer and Harko~\cite{Boehmer:2007az} who reported that the Newtonian approximation of the DM-induced deflection angle is about $\sim 77\%$ weaker than predictions from the Einstein cluster description of DM in the constant velocity region of galaxies. Kouniatalis \textit{et al.}~\cite{Kouniatalis:2025itj} recently employed the Einstein cluster to investigate its impact on light deflection near the BH; in contrast, our work focuses on directly measurable lensing quantities, as outlined above, as well as comparing two different DM descriptions and energy density distributions profiles to assess their impact on the resulting spacetime geometry.

To proceed, we adopt the line element
\begin{gather}
ds^2 = \underbrace{-f e^{\Phi}}_{g_{tt}} dt^2
+ \underbrace{f^{-1}}_{g_{rr}} dr^2
+ r^2 d\Omega^2,
\label{eq. line element metric ansatz}
\end{gather}
where $f \equiv f(r) = 1 - 2m(r)/r$. The function $m(r)$ is the mass function, while $\Phi \equiv \Phi(r)$ is the \textit{shift function}. Throughout this work, we use geometrized units with $G = c = 1$.

\section{(Dark) matter description}
\label{sec. DM dist}

When modeling the DM distribution as an anisotropic fluid with $T_{\mu}^{\nu} = \text{diag}(-\epsilon, p, p_t, p_t)$, there is no general consensus on the appropriate equation of state (EoS) for the radial pressure $p$. A number of studies adopt, for simplicity, the implicit choice $p = -\epsilon$~\cite{Jusufi:2019ltj,Toshmatov:2025rln,Xu:2018wow,Konoplya:2025ect,Hou:2018bar,Konoplya:2019sns,Jusufi:2020cpn,Liu:2021xfb,Xu:2017bpz,Liu:2024qso,He:2024amh,Myung:2025cxu,Ali:2025ney,Ma:2022jsy}, leading to the energy–momentum tensor
\begin{equation}
T_{\mu}^{\nu(V)} = \text{diag}(-\epsilon,-\epsilon,p_t,p_t).
\end{equation}
We refer to this particular EoS as \textit{vacuum} DM, since it is equivalent to a de Sitter–type EoS. This choice is, in fact, a consequence of the ansatz~\eqref{gtt}. It has been argued that this condition yields similar physical effects to more general cases where $-g_{tt} \neq g_{rr}^{-1}$ in the presence of DM~\cite{Xu:2018wow}. Moreover, this ansatz provides a straightforward, widely used procedure for constructing rotating BH solutions via the Newman–Janis algorithm~\cite{Newman:1965tw} applied to the seed metric~\cite{Xu:2018wow,Jusufi:2020cpn,Xu:2017bpz,Liu:2024qso,He:2024amh}; however, this algorithm was criticized in Ref.~\cite{Hansen:2013owa}, as it introduces pathologies and inconsistencies into the final solutions (see also Refs.~\cite{Gurses:1975vu,Beltracchi:2021ris,Beltracchi:2021tcx,Kocherlakota:2024sxx}).

On the other hand, more recent studies employ a zero radial pressure, such that the energy–momentum tensor of the DM distribution becomes
\begin{equation}
T_{\mu}^{\nu(EC)} = \mathrm{diag}(-\epsilon,0,p_t,p_t).
\end{equation}
This DM model is known as the \textit{Einstein cluster} (EC), which describes a self-gravitating matter distribution composed of particles on stable circular orbits around the center~\cite{Boehmer:2007az}, as originally proposed by Einstein~\cite{Einstein:1939ms}. The system is stabilized by the balance between gravity and the centrifugal forces from the orbiting matter, resulting in an anisotropic energy–momentum tensor with zero radial pressure. The (numerical) rotating BH immersed in an EC DM solution appeared only recently in Ref.~\cite{Fernandes:2025osu}. Moreover, a recent study suggests that EC DM arises as a solution to a non-minimally coupled vector field interacting with gravity~\cite{Fernandes:2025lon}, indicating it may effectively behave as a vector field.

We conclude our brief comparison of the two DM models here. As shown in the following section, this seemingly minor difference gives rise to non-negligible  modifications of the spacetime geometry in the near-BH region.

We now turn to the profile shape of the energy density $\epsilon(r)$, for which most studies adopt semi-analytic distribution profiles obtained from, e.g., Refs.~\cite{Hernquist:1990be,Dehnen:1993uh,Navarro:1996gj,Einasto:1996}, that effectively explain galaxy rotation curves. In this work, we employ the general form as in Ref.~\cite{Figueiredo:2023gas}
\begin{equation}
\epsilon(r)=\epsilon_0\left(\frac{r}{a_0}\right)^{-\gamma}
\left[1+\left(\frac{r}{a_0}\right)^\sigma\right]^{\frac{\gamma-\beta}{\sigma}},
\label{eq. dark matter profile}
\end{equation}
where $a_0$ and $\epsilon_0$ are constants that characterize the core size and the effective central ``density," respectively. The remaining parameters $(\sigma,\beta,\gamma)$ determine the overall shape of the distribution and are model dependent. For instance, the Hernquist~\cite{Hernquist:1990be}, Navarro–Frenk–White (NFW)~\cite{Navarro:1996gj}, and Dehnen~\cite{Dehnen:1993uh} profiles are obtained by assigning $(1,4,1)$, $(1,3,1)$, and $(1,4,\gamma_D)$, respectively, where $\gamma_D\in[0,3)$. One may directly notice that the Dehnen profile is a generalization of the Hernquist profile.

Earlier studies have reported the vanishing DM energy density near the BH horizon, both by Newtonian-based analysis~\cite{Gondolo:1999ef} and general relativistic treatment~\cite{Sadeghian:2013laa}. Analytic approximations to this effect have been constructed by incorporating a cut-off factor into the energy density profile. For instance, the authors of Refs.~\cite{Cardoso:2021wlq,Speeney:2024mas,Figueiredo:2023gas} use
\begin{equation}
\epsilon(r)\to\epsilon(r)\left(1-\frac{r_c}{r}\right),
\label{eq. edm cut off}
\end{equation}
where $r_c$ denotes the cut-off radius at which the DM density vanishes. We refer to Eq.~\eqref{eq. edm cut off} as the \textit{modified} energy density profile. In Ref.~\cite{Cardoso:2021wlq}, the authors assumed $r_c$ to coincide with the BH horizon ($r_c = 2M_{BH}$). However, this assumption may lead to violations of the energy conditions and of the transverse speed of sound~\cite{Datta:2023zmd}, motivating Ref.~\cite{Speeney:2024mas} to adopt a larger value of $r_c$\textemdash specifically $r_c = 4M_{BH}$, following the relativistic treatment results of Ref.~\cite{Sadeghian:2013laa}.

In the following discussion, we select two representative models to describe the DM energy density. We strictly employ the generalized modified form given in Eqs.~\eqref{eq. dark matter profile} and~\eqref{eq. edm cut off}. In particular, we choose the Hernquist profile and the Dehnen profile with $\gamma_{D}=3/2$. Both profiles share the feature that the enclosed DM mass converges at large distances. This is in contrast to the NFW profile, where additional outer cut-off must be introduced to avoid a divergent total mass.

\section{Spacetime geometry}
\label{sec. spacetime geometry}

Having specified the matter distribution around the BH, the spacetime metric can be obtained by solving the Einstein field equations. The radial component is determined directly from the mass function $m(r)$, which represents the combined contribution of the BH and DM:
\begin{gather}
m(r) = M_{BH} + m_{DM}(r), \\
m_{DM}(r) = \int_{r_c}^{r} 4\pi \tilde{r}^2\epsilon(\tilde{r}) d\tilde{r}. \label{eq. int mass}
\end{gather}
The total Arnowitt–Deser–Misner (ADM) mass of the system is then $M_{ADM} = M_{ BH} + M_{DM}$, where $M_{DM} = \lim_{r \to \infty} m_{DM}(r)$. The derivation and full analytic expressions of the mass functions corresponding to the modified Hernquist and Dehnen–$3/2$ profiles are provided in the \supplmat.

Throughout our analysis, we consider two representative values for both the total DM mass and the core size, each expressed relative to the BH mass: $M_{DM}/M_{BH} \in \{10,100\}$ and $a_0/M_{BH} \in \{10^3,10^4\}$. These choices are sufficient for our purposes, as they represent both relatively diluted and compact DM halos. The ratio $a_0/M_{DM}$ roughly characterizes the compactness of the DM distribution.

Although we mentioned that a general relativistic treatment indicates $r_c = 4M_{BH}$, we instead adopt the same cut-off radius $r_c = 2M_{BH}$ as in Ref.~\cite{Cardoso:2021wlq}. This choice ensures that the DM effect extends into the vicinity of the BH horizon, as we aim to identify the most significant discrepancies between the two DM descriptions in the strong-field limit.

\begin{figure}[t!]
    \centering
    \includegraphics[width=0.47\textwidth]{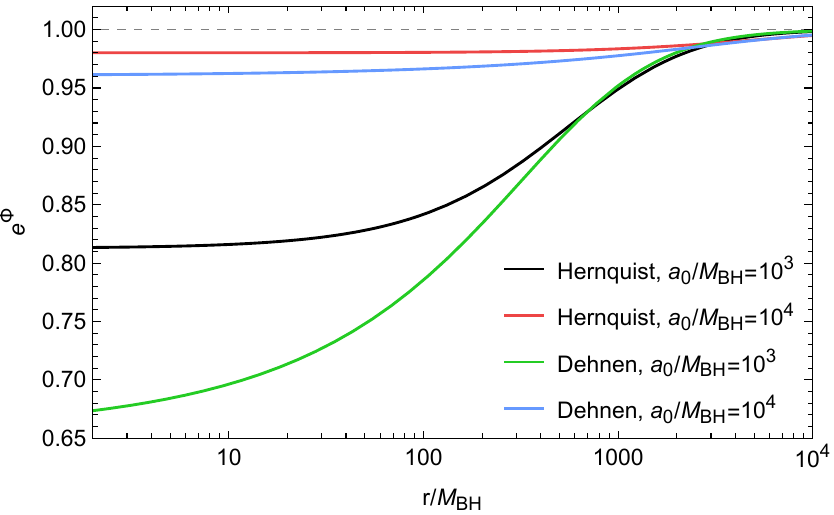}
    \caption{The shift function for the EC DM with $M_{DM}=100M_{BH}$ for the Hernquist and Dehnen-3/2 energy density profile. The gray dashed line represents the vacuum DM case, for which $\Phi(r)=0$ everywhere. The Dehnen-3/2 distribution results in a substantially higher redshift than the Hernquist profile near the BH horizon. In general, for a generalized Dehnen profile, the shift function near the BH horizon decreases as $\gamma_{D}$ increases.}
    \label{fig. shift function}
\end{figure}

Now we are left with the time component, in particular the shift function $\Phi(r)$. Following the formalism in Ref.~\cite{Fauzi:2025rcc}, the shift function for each DM case satisfies (see the \supplmat)
\begin{equation}
\frac{d\Phi^{(V)}}{dr}=0,\qquad \frac{d\Phi^{(EC)}}{dr}=\frac{8\pi \epsilon(r) r^2}{r-2m(r)}.
\label{eq. shift function}
\end{equation}
In an asymptotically flat spacetime, one requires the boundary condition $\Phi(r)|_{r\to\infty}=0$. This implies that EC DM introduces an additional redshift factor, since the shift function shall increase towards larger $r$. Meanwhile, vacuum DM simply leads to $-g_{tt}=g_{rr}^{-1}$. Additionally, Ref.~\cite{Cardoso:2021wlq} has obtained the analytic form of the $-g_{tt}$ component for the modified Hernquist profile with $r_c = 2M_{BH}$. Other approximate analytical solutions for generalized DM energy density profiles of Eq.~\eqref{eq. dark matter profile} in the EC description have been obtained in Ref.~\cite{Konoplya:2022hbl}.

We compare the non-zero shift function induced by the EC DM for both the Hernquist and Dehnen-3/2 profiles in Fig.~\ref{fig. shift function}, where we set $M_{DM}/M_{BH}=100$ for better visualization. We obtain the shift function by numerically integrating Eq.~\eqref{eq. shift function} backward from (numerical) infinity, imposing the aforementioned boundary condition for an asymptotically flat spacetime. The plot illustrates that the additional redshift factor induced by the EC DM remains nonzero for radii $r \gtrsim a_0$, although the main deviation from the zero shift function occurs predominantly at scales of the order of $a_0$ or smaller. A more compact configuration (smaller $a_0/M_{\rm BH}$) leads to a larger deviation from the zero shift function over a shorter radial range. Moreover, the Dehnen-3/2 profile produces a substantially stronger redshift factor near the BH horizon compared to the Hernquist profile, indicating that the discrepancy between EC DM and vacuum DM is likely model dependent. As we will show, this additional redshift factor plays a crucial role in electromagnetic observations, and its impact becomes non-negligible in the strong-field regime.

\section{Resulting observables}
\label{sec. electromagnetic obs}

Since we are dealing with a static and spherically symmetric spacetime, the geodesic equations for photons take a simple form. We provide the expression of the geodesic equation and its derivation in the \supplmat. The corresponding photon effective potential takes the form
\begin{equation}
V(r)=\frac{-g_{tt}}{r^2}=\left[1-\frac{2m(r)}{r}\right]\frac{e^{\Phi}}{r^2}.
\end{equation}
This potential determines the turning point of photons approaching the system. Since its form depends explicitly on the shift function, we expect the EC DM scenario to yield electromagnetic observables that differ from those obtained under the vacuum DM approximation.

\begin{figure}[t!]
    \centering
    \includegraphics[width=0.48\textwidth]{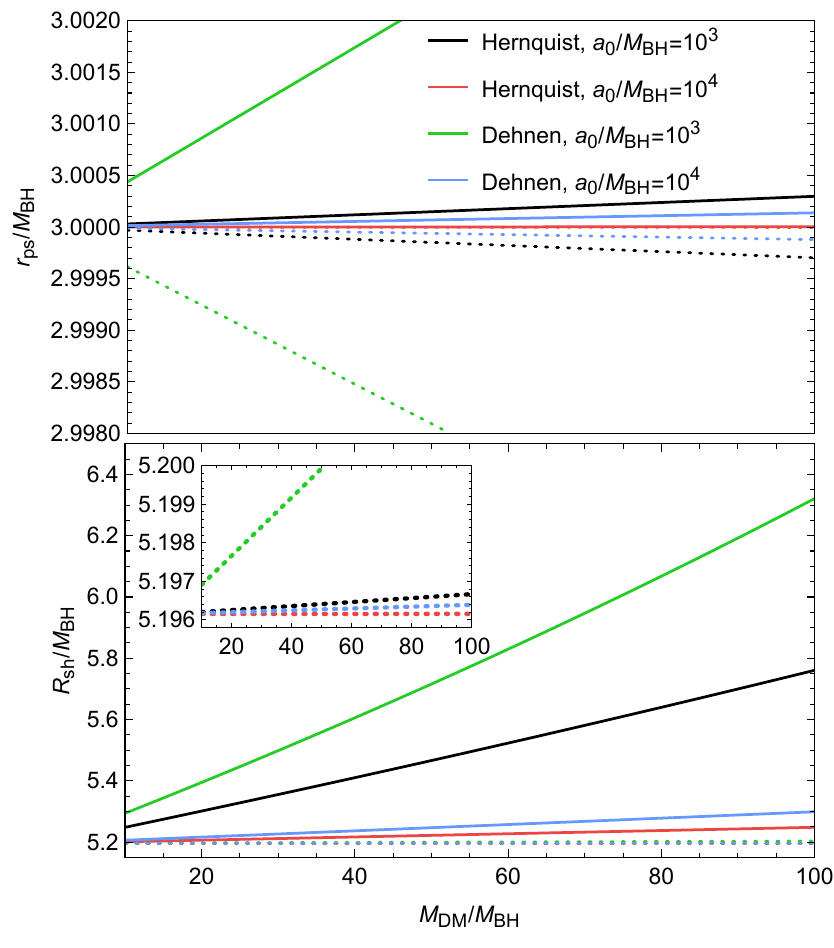}
    \caption{(Top) Photon sphere radius and (Bottom) shadow radius for the EC DM and vacuum DM as functions of the DM mass, shown by solid and dotted lines, respectively. The photon sphere radius behaves in opposite ways in the EC and vacuum DM cases. The BH shadow radius is significantly more sensitive to the DM mass in the EC DM scenario than in the vacuum DM case; a closer look at the vacuum DM shows that the shadow radius barely deviates from the Schwarzschild BH value of $3\sqrt{3}M_{BH}\approx5.196M_{BH}$.}
    \label{fig. rps rsh}
\end{figure}

\textbf{\textit{Photon sphere and shadow}}\textemdash To calculate the BH shadow radius, one first needs to determine the location of the photon sphere $r_{ps}$. This is obtained by extremizing the effective potential, i.e., by solving $\partial_{r}V|_{r=r_{ps}}=0$. Since the $\Phi(r)$ itself is obtained numerically, we solve the $r_{ps}$ by a generic numerical root-finding method. Once the photon sphere is located, the critical impact parameter $b_c$ follows from the relation $b_c = V(r_{ps})^{-1/2}$. Since the spacetime is asymptotically flat, the shadow radius $R_{sh}$ as seen by a distant observer is given by~\cite{Perlick:2021aok}
\begin{equation}
R_{sh}\approx b_c.
\label{eq. rsh rps}
\end{equation}

The resulting photon sphere radius, together with the BH shadow radius, is shown in Fig.~\ref{fig. rps rsh}. One can clearly observe that the photon sphere exhibits opposite trends in the two DM models: increasing the total DM mass shifts the photon sphere outward in the EC DM case, whereas in the vacuum DM case, it is shifted inward. The latter behavior commonly appears in models that adopt a de Sitter–like vacuum equation of state to describe the effective matter sourcing the geometry, as is typical in regular BH solutions (see, e.g., Refs .~\cite {Gera:2024qob,Carballo-Rubio:2022nuj}). Nonetheless, the deviation of the photon sphere from its Schwarzschild value ($r_{ps}=3M_{BH}$) remains extremely small, even in the most compact configurations with small $a_0/M_{DM}$.

On the other hand, a significant discrepancy arises in the BH shadow radius. Specifically, in the EC DM model, increasing the DM mass results in a much steeper increase in the shadow radius than in the vacuum dark matter case. This behavior is directly linked to the additional redshift factor sourced by EC DM: it suppresses the $g_{tt}$, thereby lowering the peak of the photon effective potential and increasing the critical impact parameter. By contrast, in the vacuum DM model, the spacetime geometry near the BH remains nearly indistinguishable from the Schwarzschild solution, since the modification enters only through the mass function and becomes important predominantly at large distances.

This point is crucial when using BH shadow observations to constrain DM parameters. For example, by simply applying the $2\sigma$ Sagittarius A* shadow radius constraint from Ref.~\cite{Vagnozzi:2022moj} (see also Ref.~\cite{EventHorizonTelescope:2022xqj}),
\begin{equation}
    4.21M_{BH}\lesssim R_{sh}^{(\text{Sgr A*})} \lesssim 5.56M_{BH},
\end{equation}
to the Hernquist distribution with $a_0/M_{BH}=10^3$, the EC DM model sets an upper bound on total DM mass at $M_{DM}/M_{BH}\approx67$. In contrast, the vacuum DM model permits a much greater DM mass, up to the threshold where the halo forms a secondary horizon. This is consistent with Ref.~\cite{Konoplya:2019sns} who found that notable changes in the BH shadow radius in vacuum DM models appear only when DM is highly concentrated near the BH, i.e., when $a_0\sim\sqrt{3M_{BH}M_{DM}}$. Thus, the DM description\textemdash whether as an Einstein cluster or a vacuum fluid\textemdash must be carefully chosen, as it yields qualitatively different astrophysical conclusions.

\begin{figure*}[t!]
    \centering
    \includegraphics[width=0.95\textwidth]{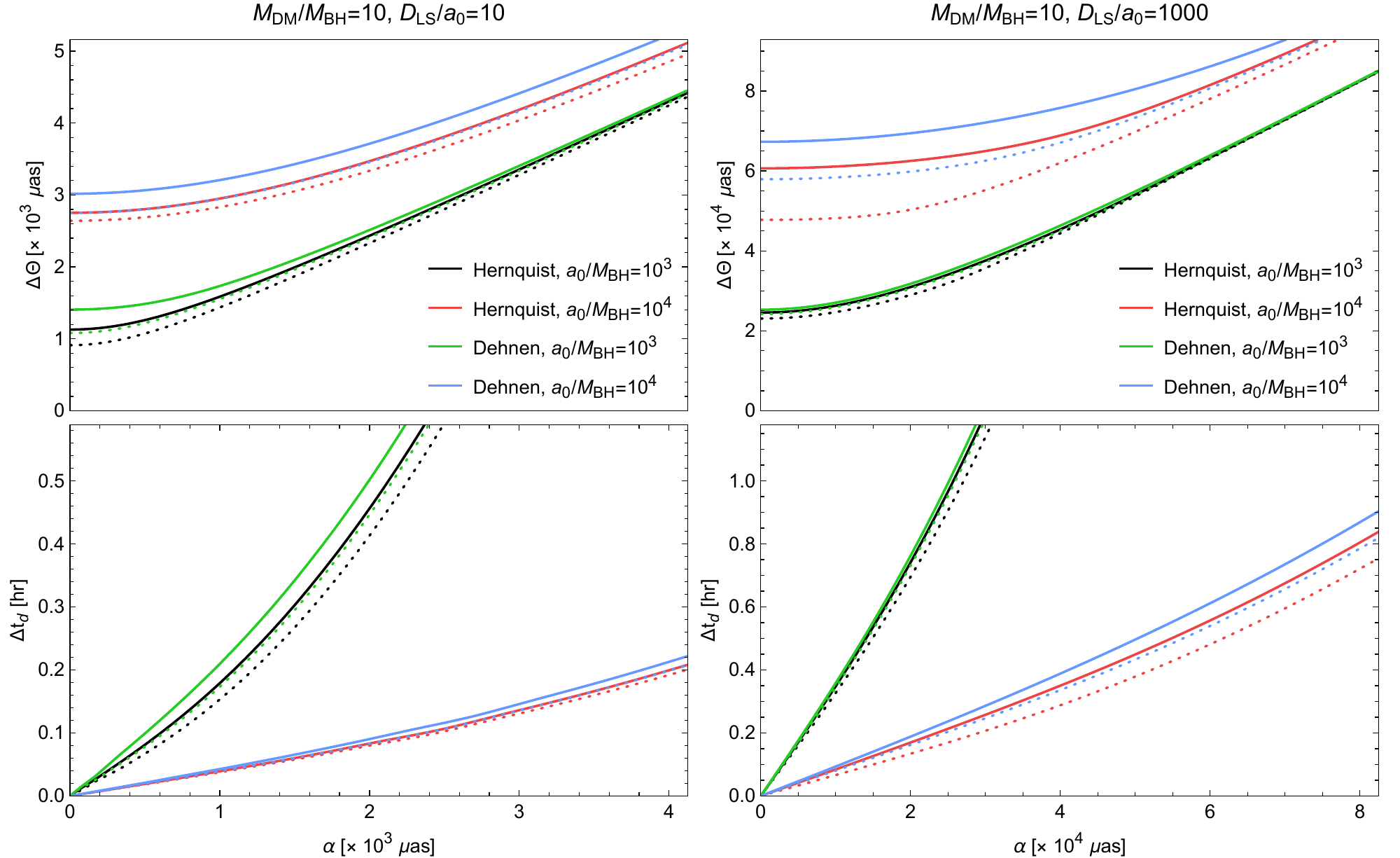}
    \caption{(Top) Separation angle between the lensed primary and secondary images as a function of $\alpha$, and (Bottom) their differential time delay, for EC DM (solid lines) and vacuum DM (dotted lines). As a reference, the shadow angular diameter of a Schwarzschild BH in this setup is $\sim 21.4\,\mu\text{as}$. Significant discrepancies in the separation angle between EC DM and vacuum DM are observed at the Einstein ring (when $\alpha = 0$). On the other hand, the differential time delay shows the largest differences at higher source inclinations.}
    \label{fig. alpha theta d time}
\end{figure*}

\begin{figure}[t!]
    \centering
    \includegraphics[width=0.48\textwidth]{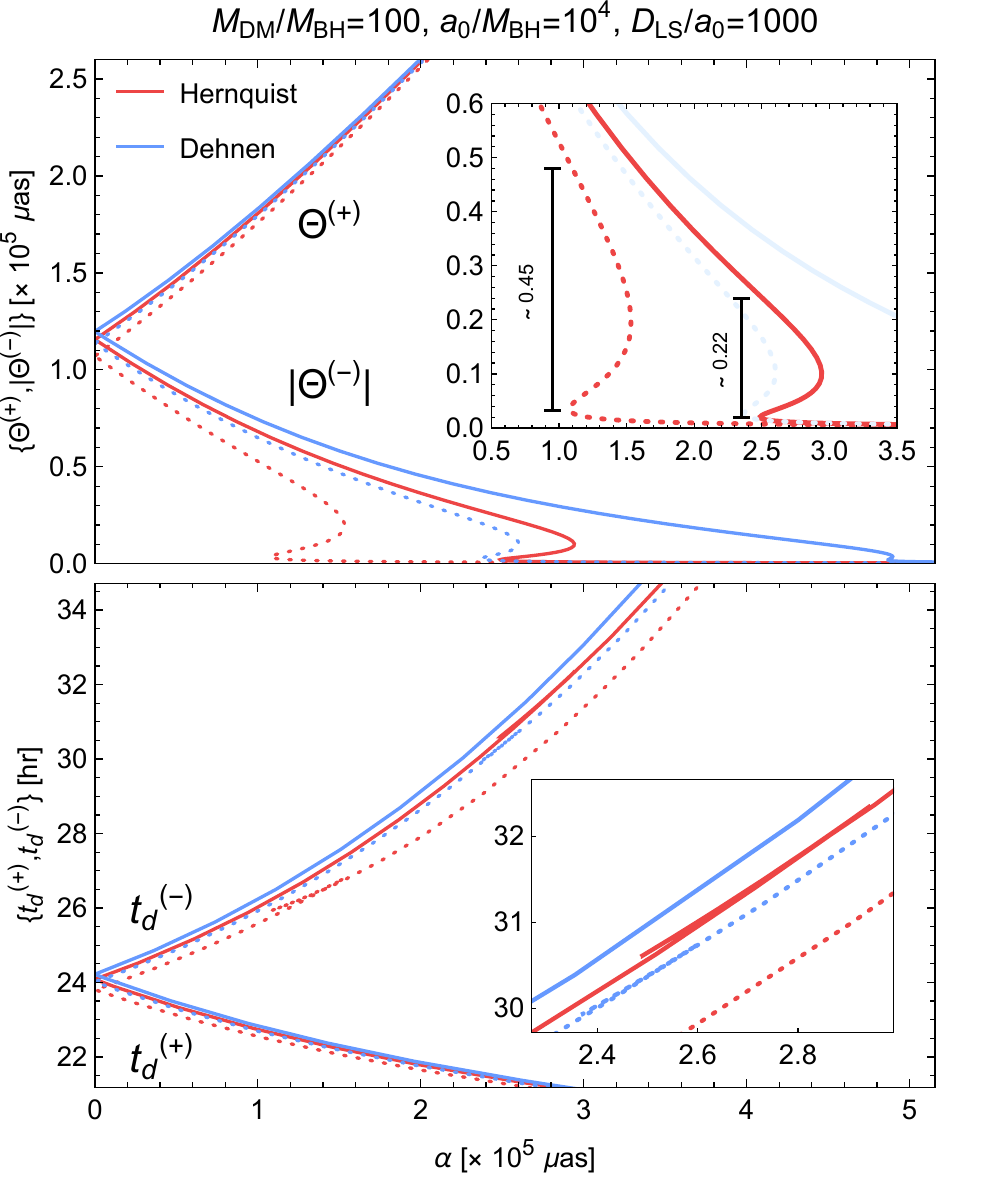}
    \caption{(Top) Angular positions of the primary and secondary images as functions of $\alpha$, and (Bottom) their corresponding time delays, for EC DM (solid lines) and vacuum DM (dotted lines), with $M_{DM}/M_{BH}=100$ and $D_{LS}/a_0=1000$. At certain source inclinations in both the EC DM and vacuum DM cases, the DM produces multiple secondary images. A closer examination of their corresponding time delays indicates that these images have similar time delays.}
    \label{fig. lensing delay}
\end{figure}

\textbf{\textit{Lensing observables}}\textemdash Next, we investigate the discrepancy between the two DM models in their lensing observables. Similar analyses of DM effects on lensing have recently been carried out in Ref.~\cite{Kouniatalis:2025itj} for the EC DM model with the modified Hernquist profile of Ref.~\cite{Cardoso:2021wlq}. Numerous studies have also explored lensing signatures in various DM scenarios (see, for example, Refs.~\cite {Jusufi:2022jxu,Pantig:2022toh,Liu:2023xtb,Molla:2025yoh,Molla:2023yxn} for some of the latest results).

The gravitational lensing setup is described as follows. A source at angular position $\alpha$ relative to the observer produces two images at angles $\Theta^{(+)}$ and $\Theta^{(-)}$, corresponding to the primary and secondary images, respectively~\cite{Virbhadra:2008ws}. The time delay for each image, $t_d^{(\pm)}$, is defined as the difference between the travel time in curved spacetime and that in flat spacetime. The observable quantities are the angular separation $\Delta\Theta = |\Theta^{(+)} - \Theta^{(-)}|$ and the differential time delay $\Delta t_d = |t_d^{(+)} - t_d^{(-)}|$. An illustration and computational details are provided in the \supplmat.

For our purposes, we assume the source lies \textit{within} or \textit{near} the DM halo at a distance $D_{LS}$ from the BH, while the observer is located at $D_{OL}$ in a region where spacetime is effectively flat ($g_{rr} \approx -g_{tt} \approx 1$). This setup ensures a significant influence of the DM-induced redshift, particularly on the primary images. If both $D_{OL}$ and $D_{LS}$ were very large, the closest approach of the primary rays would occur outside the DM halo, where the spacetime is approximately Schwarzschild and the redshift effect becomes negligible. We set $D_{OL}/M_{BH} \approx 10^{11}$ and choose the $D_{LS}$ relative to the DM core size $a_0$ as $D_{LS}/a_0 \in \{10,1000\}$.

For reference, while adopting the mass of Sagittarius A* ($\sim 4.3\times 10^6 M_{\odot}$~\cite{GRAVITY:2023avo}) for $M_{BH}$ in our settings, we obtain $D_{OL}\approx 20\,\text{kpc}$, $D_{LS}\in (0.002,2)\,\text{pc}$, and $a_0 \approx\{2\times10^{-4},2\times10^{-3}\} \,\text{pc}$. Obviously, these choices are not intended to represent the realistic astrophysical environment of Sagittarius A*. We instead adopt this configuration to illustrate the qualitative behavior of both models and, to avoid confusion, use the mass of Sagittarius A* to fix the units of our results.

We compare the EC and vacuum DM predictions for both observables in Fig.~\ref{fig. alpha theta d time}, focusing on the case with $M_{DM}/M_{BH}=10$. In all configurations, EC DM produces stronger lensing effect than vacuum DM for both image separation and differential time delay. The largest differences in the image position between the vacuum DM and EC DM cases appear when the source, lens, and BH are aligned ($\alpha = 0$), resulting in the so-called Einstein ring~\cite{Wambsganss:1998gg}. The size of this discrepancy depends on both the source distance $D_{LS}$ and the DM core size. A smaller core causes more significant differences for sources near the BH, while a larger core amplifies the effect for more distant sources. 

The smallest and largest differences in the angular position of the Einstein ring between the EC DM and vacuum DM cases in our comparison occur at $D_{LS}/a_0 = 1000$. The smallest difference appears for the more compact DM configuration described by the Dehnen-3/2 profile ($2.52\times10^4\,\mu\text{as}$ for EC DM and $2.42\times10^4\,\mu\text{as}$ for vacuum DM), while the largest occurs for the less compact DM configuration described by the Hernquist profile ($6.06\times10^4\,\mu\text{as}$ for EC DM and $4.78\times10^4\,\mu\text{as}$ for vacuum DM). Therefore, our results suggest that EC DM predicts an Einstein ring diameter that is roughly $\sim 4\%-27\%$ larger than that predicted by vacuum DM.

Our results also show that the discrepancy in the differential time delay between the vacuum and EC DM cases is insignificant given the current observational capabilities. We predict a discrepancy on a timescale of less than an hour; even the $\sim 2-3$ hour delay reported for the Einstein Cross Q2237+0305 remains unreliable, as it is associated with relatively large uncertainties~\cite{Dai:2003ie,Fedorova:2026fet}. This indicates that we may not be able to distinguish between the vacuum and EC DM models based on the observed differential time delay alone.

In the setup with $M_{DM}/M_{BH}=100$ and $D_{LS}/a_0=10^3$, in Fig.~\ref{fig. lensing delay}, we find an interesting phenomenon arising from the DM distribution: there exists a range of $\alpha$ that produces multiple secondary images across all DM models. This phenomenon was also reported in a recent lensing study of EC DM around a BH~\cite{Kouniatalis:2025itj} on the deflection angle. A similar feature is clearly visible in Fig.~4c of Ref.~\cite{Xavier:2023exm}; however, the multiplication of secondary images in our case occurs for a significantly less compact DM configuration, at angular scales several orders of magnitude larger than the shadow boundary. Our analysis, supplemented by the results of Ref.~\cite{Kouniatalis:2025itj}, shows that the occurrence of multiple secondary images is a consequence of a range of impact parameters that yield roughly the same deflection angle (see the \supplmat).

The discrepancy in multiple secondary images between the two DM descriptions is reflected in their positions and separations. The vacuum DM case produces multiple lensed images at smaller source inclinations as seen by the observer, but with an overall larger separation than in the EC DM case: for the Hernquist profile DM, the vacuum DM and EC DM scenarios yield the largest separation of the multiple secondary images of approximately $\sim 0.45\times10^{5}\,\mu\text{as}$ and $\sim 0.22\times10^{5}\,\mu\text{as}$, respectively (see the top inset of Fig.~\ref{fig. lensing delay}). However, the time delays associated with these multiple lensed images are roughly the same, implying that all events in these lensed images occurred at the same time as observed by a distant observer. This may become an important potential signature of compact DMs around a BH, regardless of the chosen description of the DM halo itself.

\section{Conclusion and discussion}
\label{sec. conclusion}

We have compared two models of anisotropic dark matter (DM) distributions around a black hole (BH): the vacuum DM model and the Einstein cluster (EC) DM model. The primary physical distinction is the presence of negative radial pressure on the vacuum DM, which influences the time component of the metric. Vacuum DM acts as a ``de Sitter" fluid and does not alter the shift function on the metric, whereas EC DM introduces an additional redshift factor whose behavior depends on the energy density profile. Consequently, the two models produce distinct signatures in electromagnetic observables. Our analysis shows that the photon sphere radius increases with DM mass in the EC model, while it shifts slightly inward for vacuum DM. Notably, EC DM exerts a significantly stronger effect on the BH shadow radius compared to vacuum DM. Lensing observables, including image positions and the differential time delay between primary and secondary images, also exhibit substantial differences between the two models.

We therefore conclude that the theoretical modeling of the DM halo surrounding a BH is crucial for determining its electromagnetic observables. If the actual DM distribution is best represented by an Einstein cluster, the vacuum DM approximation is insufficient for probing the geometry near the BH. Combining shadow measurements with strong-field lensing signatures offers a promising approach to distinguish between different DM models, particularly when supplemented by independent data from the dynamics of nearby objects\textemdash especially stars orbiting the BH, such as the S-stars around the Sagittarius A* BH~\cite{Hees:2017aal}. Future observational data may enable the exclusion or confirmation of competing DM descriptions.

 Several caveats warrant attention. The adopted setup and assumptions may not fully capture a realistic cosmological settings. For example, time delay analyses in large scale gravitational lensing (e.g., HE 0435-1223~\cite{H0LiCOW:2016qrm}) typically include cosmic expansion to infer the Hubble constant, which was not addressed in this study. Furthermore, lensing observables, especially time delays, are sensitive to perturbing matter both near the lens and along the line of sight~\cite{Seljak:1994wa}. Future work that incorporates the cosmological constant into the Einstein field equations and accounts for perturbers from nearby galaxies and the environment along the light paths would be valuable.

Other prospective directions may also be considered. The distinct effective potentials in the EC DM and vacuum DM cases indicates that they may lead to differences in their gravitational-wave (GW) signatures, e.g., in their quasinormal modes and GW lensing echoes~\cite{Jusufi:2019ltj,Toshmatov:2025rln,Cardoso:2021wlq,Kouniatalis:2025itj} (see, e.g., Refs.~\cite{Ezquiaga:2020dao,Gondan:2021fpr} for relevant discussions of GW lensing echoes). In addition, the appearance of the accretion disk around the BH may also be relevant, as we expect that EC DM and vacuum DM would yield distinct innermost circular orbits and different redshift effects on the disk intensity (see, e.g., Ref.~\cite{Macedo:2024qky}). Furthermore, one could perform a similar analysis to that presented here to compare EC DM and vacuum DM observables in their rotating configurations (e.g., those obtained numerically~\cite{Fernandes:2025osu} and those derived via the Newman–Janis algorithm~\cite{Xu:2018wow,Jusufi:2020cpn,Xu:2017bpz,Liu:2024qso,He:2024amh}), since realistic BHs are expected to be non-static.

Nevertheless, our results capture robust features of strong-field lensing around BHs embedded in DM and remain relevant for future theoretical and observational investigations.

\section*{Acknowledgements}

We thank Byon Jayawiguna, Faris Darmawan, and Naufal Athaullah for the valuable discussions on the early stage of this work. We also thank the anonymous referee for the valuable suggestions on this manuscript. HSR is supported by Hibah PUTI Q1 UI No.~PKS-196/UN2.RST/HKP.05.00/2025.

\begin{appendix}
\end{appendix}

\onecolumngrid
\newpage
\pagenumbering{gobble}

\renewcommand{\theequation}{S.\arabic{equation}}
\renewcommand{\thefigure}{S.\arabic{figure}}
\setcounter{equation}{0}
\setcounter{figure}{0}

\section*{Supplementary Material}
\vspace{-0.3cm}
\noindent\rule{\linewidth}{0.5pt}
\vspace{-1cm}
\label{sec. supplementary material}

\subsection*{Deriving the metric components}

For a perfect fluid description of the energy–momentum tensor in a static and spherically symmetric spacetime, the metric components can be obtained by solving the Einstein field equations $G_{\mu\nu}=8\pi T_{\mu\nu}$. In our case, the source term $T_{\mu\nu}$ is provided by the DM. We focus on the two relevant equations: $G_0^0=8\pi T_0^0$ and $G_1^1=8\pi T_1^1$. Using the line element given in Eq.~\eqref{eq. line element metric ansatz}, we obtain
\begin{align}
    \frac{f'}{r} +\frac{f}{r^2} - \frac{1}{r^2} &= -8\pi \epsilon, & (G_0^0=8\pi T_0^0)
	\label{eq. 00 Guv = Tuv}\\
	\frac{f\Phi'}{r}+\frac{f'}{r}+\frac{f}{r^2}-\frac{1}{r^2} &= 8\pi p, \label{eq. 11 Guv = Tuv}& (G_1^1=8\pi T_1^1)
\end{align}
where $'$ denotes differentiation with respect to $r$. Equation~\eqref{eq. 00 Guv = Tuv} can be straightforwardly integrated, yielding
\begin{equation}
    f=1-\frac{2C}{r}-\frac{2m_{DM}(r)}{r},\qquad m_{DM}(r)=\int^r_{r_c} 4\pi\tilde{r}^2\epsilon(\tilde{r})d\tilde{r},
    \label{eq. f and m}
\end{equation}
where $C$ is an integration constant and $m_{DM}(r)$ is the Misner–Sharp mass of the DM. One readily concludes that $C = M_{BH}$ by requiring a Schwarzschild geometry in vacuum ($\epsilon \to 0$). In our DM setup, since there is no DM mass contribution inside the cut-off radius $r_c$, the integration of the mass function begins at $r_c$. The function $f(r)$ then reads
\begin{equation}
    f(r)=1-\frac{2m(r)}{r},\qquad m(r)=M_{BH}+m_{DM}(r)
    \label{eq. f and m final}
\end{equation}
with $m(r)$ is the total Misner-Sharp mass of the system.

The resulting mass functions for the modified Hernquist and Dehnen-3/2 energy density profiles used in this work can be computed analytically. Integrating Eq.~\eqref{eq. f and m} for both modified density profiles yields
\begin{align}
    m_{DM}^{(H)}(r)=&\frac{2\pi a_0^4 \epsilon_0}{(a_0+r_c)}\frac{(r-r_c)^2}{(r+a_0)^2}\\
    m_{DM}^{(D)}(r)=&\frac{8}{3}\pi a_0 \epsilon_0\left[\frac{a_0\sqrt{r}[a_0(r-3r_c)-2r_cr]}{(a_0+r)^{3/2}}+\eta\right]\\
    \eta=&\frac{2a_0r_c^{3/2}}{\sqrt{a_0+r_c}}\notag
\end{align}
where the superscripts $(H)$ and $(D)$ denote the Hernquist and Dehnen-3/2 mass profiles, respectively. Taking the limit $r \to \infty$ of the mass function yields the Arnowitt–Deser–Misner (ADM) mass of the DM ($M_{DM}$),
\begin{gather}
    \lim_{r\to\infty}m_{DM}^{(H)}(r)=\frac{2\pi a_0^4\epsilon_0}{a_0+r_c}\equiv M_{DM}^{(H)},\\
    \lim_{r\to\infty}m_{DM}^{(D)}(r)=\frac{8}{3}\pi a_0\xi\epsilon_0 \equiv M_{DM}^{(H)},\\
    \xi=a_0(a_0-2r_c)+\eta.\notag
\end{gather}
Therefore, expressing the result in terms of the total DM mass and dropping the superscripts, the mass function reads
\begin{align}
m_{DM}^{(H)}(r)=&M_{DM}\frac{(r-r_c)^2}{(r+a_0)^2}\\
m_{DM}^{(D)}(r)=&M_{DM}\xi^{-1}\left[\frac{a_0\sqrt{r}[a_0(r-3r_c)-2r_cr]}{(a_0+r)^{3/2}}+\eta\right]\\
\end{align}

We are now left with the $-g_{tt}$ component of the metric, specifically the shift function $\Phi$. Following Ref.~\cite{Fauzi:2025rcc}, we adopt the following ansatz for the equation of state:
\begin{equation}
    p(r)=-\epsilon(r)[1-F(r)]
    \label{eq. eos ansatz}
\end{equation}
where $F$ is an arbitrary function. This ansatz is particularly useful for studying deviations from the (de Sitter) vacuum equation of state, for which $F=0$. In our case, we simply take $F=1$ for the EC DM and $F=0$ for the vacuum DM, so that $F(r)\to F$ is constant. Using the ansatz~\eqref{eq. eos ansatz}, substituting Eq.~\eqref{eq. 00 Guv = Tuv} into Eq.~\eqref{eq. 11 Guv = Tuv}, and rearranging, we obtain
\begin{equation}
    \frac{f\Phi'}{r} = 8\pi\epsilon F.
    \label{eq. G11 p ansatz}
\end{equation}
Using $f$ from Eq.~\eqref{eq. f and m final}, we finally obtain
\begin{equation}
    \frac{d\Phi}{dr} = \frac{8\pi r^2 \epsilon(r) F}{r-2m(r)}.
\end{equation}
Therefore, for EC DM with $F=1$ and vacuum DM with $F=0$, we obtain Eq.~\eqref{eq. shift function}, which describes the behavior of the shift function in both the EC DM and vacuum DM cases.

\subsection*{Deriving the geodesic equation for photon}

The derivation of the geodesic equations in Eqs.~\eqref{eq. dphi dr} and~\eqref{eq. dt dr} is straightforward. We begin with the general geodesic equation
\begin{equation}
\frac{d^2x^\mu}{d\tau^2}+\Gamma_{\alpha\beta}^\mu \frac{dx^\alpha}{d\tau}\frac{dx^\beta}{d\tau}=0,
\end{equation}
with $x^{\mu}=(t,r,\theta,\phi)$ and $\tau$ an affine parameter. Restricting the motion to the equatorial plane ($\theta=\pi/2$, $\dot{\theta}=0$), we obtain the following constants of motion:
\begin{equation}
-g_{tt}\dot{t}=E,\qquad r^2\dot{\phi}=L,
\label{eq. const of motion}
\end{equation}
where the dotted symbols denote derivatives with respect to the affine parameter $\tau$. The four-velocity condition for null particles reads
\begin{equation}
g_{tt}\dot{t}^2+g_{rr}\dot{r}^2+r^2\dot{\theta}^2+r^2\sin\theta\dot{\phi}^2=0.
\end{equation}
Rearranging the expression and substituting Eq.~\eqref{eq. const of motion} yields
\begin{equation}
-g_{tt}g_{rr}\left(\frac{dr}{d\tau}\right)^2+V(r)L^2=E^2,\qquad V(r)=\frac{-g_{tt}}{r^2}=\left[1-\frac{2m(r)}{r}\right]\frac{e^{\Phi}}{r^2},
\label{eq. geodesic T+V=E}
\end{equation}
where $V(r)$ is the photon effective potential. Applying the chain rule to Eq.~\eqref{eq. geodesic T+V=E}, we finally obtain
\begin{align}
\frac{d\phi}{dr}&=\frac{d\phi}{d\tau}\left(\frac{dr}{d\tau}\right)^{-1}=\pm\frac{1}{r^2} \left[\frac{-g_{tt}g_{rr}}{b^{-2} - V(r)}\right]^{1/2},\label{eq. dphi dr}\\
\frac{dt}{dr}&=\frac{dt}{d\tau}\left(\frac{dr}{d\tau}\right)^{-1}=\pm\frac{1}{b} \left[\frac{-g_{tt}g_{rr}}{b^{-2} - V(r)}\right]^{1/2},\label{eq. dt dr}
\end{align}
with $b \equiv L/E$ is known as the impact parameter. These two equations are then used to compute the photon trajectories in the spacetime, particularly for determining the lensing observables.

\subsection*{Computing the lensing observables}

\begin{figure*}[htbp!]
    \centering
    \includegraphics[width=1\textwidth]{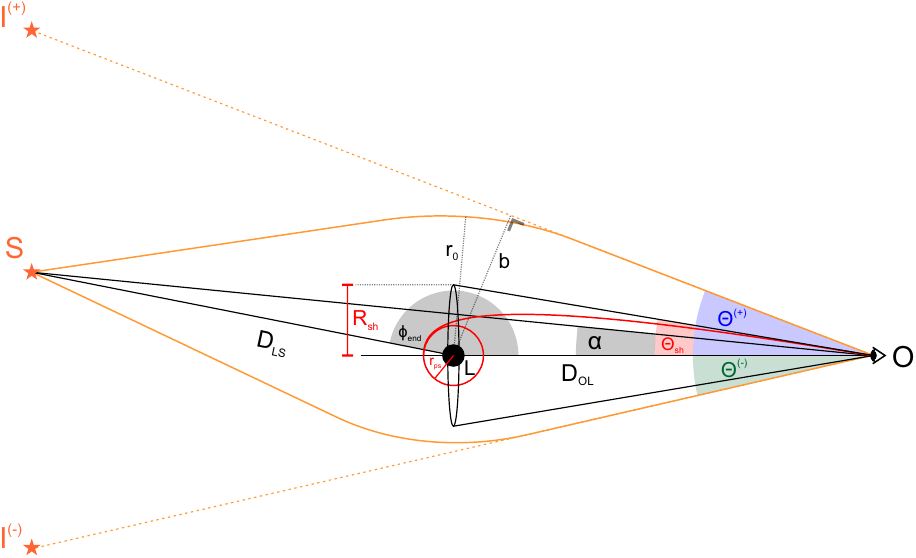}
    \caption{Illustration of the BH lensing setup.}
    \label{fig. lensing illus}
\end{figure*}

Solving the geodesic equation to compute the lensing observables fully numerically for our case requires several steps. An illustration of the corresponding lensing configuration is shown in Fig.~\ref{fig. lensing illus}. 

The lensing observables are described as follows. Two images\textemdash namely, the primary image $I^{(+)}$ at an angle $\Theta^{(+)}$ and the secondary image $I^{(-)}$ at an angle $\Theta^{(-)}$\textemdash of a light source located at a distance $D_{LS}$ from the BH and at an angle $\alpha$ with respect to the observer are observed on opposite sides of the BH. The light rays producing both images travel longer paths than they would in flat spacetime and, further affected by the redshift factor, require a longer time to reach the observer. This results in the time delays $t_d^{(\pm)}$ for the two images. 

The quantities of interest for observation are then the angular image positions $\Theta^{(\pm)}$ and the time delays $t_d^{(\pm)}$ as functions of $\alpha$. All of these quantities can be obtained by solving the null geodesic equations given in Eqs.~\eqref{eq. dphi dr} and~\eqref{eq. dt dr}.

The integration of the geodesic equation in Eq.~\eqref{eq. dphi dr} begins from the observer at $D_{OL}$ to the closest approach distance $r_0$\textemdash also known as the turning point, where $V(r_0)=b^{-2}$\textemdash and is then continued to the source location at $D_{LS}$. The integration yields an angle $\phi_{end}$ (see Fig.~\ref{fig. lensing illus}),
\begin{equation}
\phi_{end}=-\int_{D_{OL}}^{r_0}\left(\frac{d\phi}{dr}\right)dr+\int_{r_0}^{D_{LS}}\left(\frac{d\phi}{dr}\right)dr.
\label{eq. phi end}
\end{equation}
Fixing the values of $M_{BH}$, $M_{DM}$, and $a_0$ gives $d\phi/dr \to d\phi(r,b)/dr$ and $r_0 \to r_0(b)$. Consequently, once $D_{OL}$ and $D_{LS}$ are fixed, the integral and $\phi_{\rm end}$ depend solely on the impact parameter, so that $\phi_{\rm end} \to \phi_{\rm end}(b)$. From this quantity, one obtains the relation between the source angle $\alpha$ and $\phi_{end}$,
\begin{equation}
\sin\alpha=\frac{D_{LS}}{d}\sin\phi_{end},
\label{eq. sin alpha phi end}
\end{equation}
where $d=\sqrt{D_{LS}^2+D_{OL}^2-2D_{LS}D_{OL}\cos\phi_{end}}$, which gives $\alpha(b)$. Meanwhile, the image angle $\Theta^{(\pm)}$ is computed straightforwardly from elementary geometry,
\begin{equation}
\Theta^{(\pm)}=\tan^{-1}\frac{b}{D_{OL}},
\label{eq. Theta pm}
\end{equation}
yielding $\Theta^{(\pm)}(b)$.

In the simple case\textemdash \textit{i.e.}, when there is no multiple value of $b$ corresponding to a single $\alpha$\textemdash the relation $\Theta^{(\pm)}(\alpha)$ can be obtained numerically via interpolation. The primary $(+)$ or secondary $(-)$ images are selected based on the sign of $\alpha$: if $\alpha>0$, the result is assigned to the primary image, whereas for $\alpha<0$ it corresponds to the secondary image. Afterward, the magnitude of $\alpha$ for the secondary image can be taken as positive. This is permissible due to spherical symmetry, which allows a $180^\circ$ rotation around the $x$-axis without changing the physical configuration.

However, the same interpolation procedure cannot be applied when there exist multiple values of $b$ for a given $\alpha$. This occurs in our configuration with $M_{DM}/M_{BH}=100$ and $D_{LS}/a_0=10^3$. In this situation, we instead present the relation $\Theta^{(\pm)}(\alpha)$ using a parametric plot, as shown in Fig.~\ref{fig. alpha theta d time}, with the absolute value of $\alpha$ placed on the $x$-axis.

Additionally, the time delay $t_d$ is computed in a similar manner. The integration of Eq.~\eqref{eq. dt dr} gives the elapsed time required for a photon emitted from the source to reach the observer. It is expressed as
\begin{equation}
t_{end}=\left|\int_{D_{OL}}^{r_0}\left(\frac{dt}{dr}\right)dr\right|+\left|\int_{r_0}^{D_{LS}}\left(\frac{dt}{dr}\right)dr\right|,
\label{eq. phi end}
\end{equation}
where the absolute values are used to avoid negative contributions, since the elapsed time in both segments must be positive. Similar to Eq.~\eqref{eq. phi end}, this gives $t_{\rm end}(b)$. The time required for a photon to travel a distance $d$ in flat spacetime is simply $t_f = cd$; since we set $c=1$, this reduces to $t_f = d$. The time delay is then defined as $t_d = t_{\rm end} - t_f$. Because the distance $d$ depends on $\phi_{end}(b)$, we have $d\to d(b)$, and the time delay can be written as
\begin{equation}
t_{d}(b)=t_{end}(b)-d(b).
\end{equation}
The same numerical procedure used to obtain $\Theta^{(\pm)}(\alpha)$ can then be applied to acquire $t_d(\alpha)$.

\subsection*{Multiplication of secondary images}

\begin{figure*}[htbp!]
    \centering
    \includegraphics[width=0.7\textwidth]{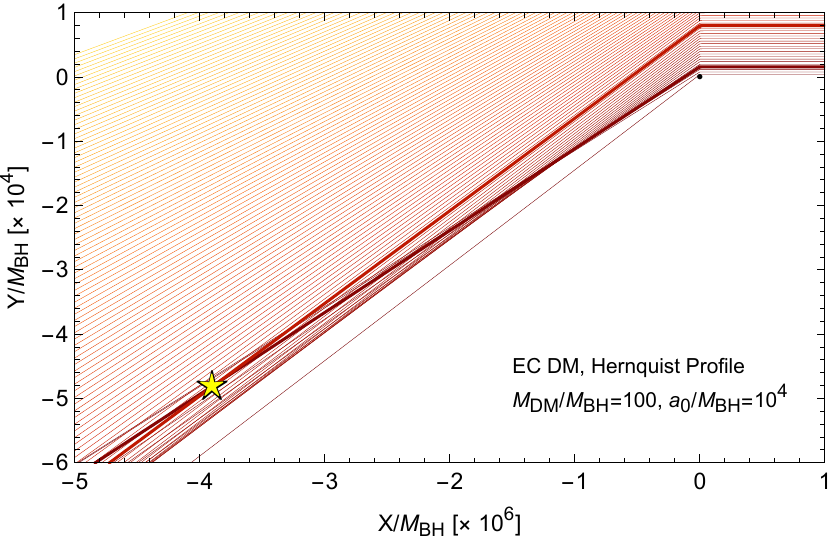}
    \caption{Photon trajectories originating from arbitrary sources on the left side of the figure, propagate toward the BH, and then continue to the observer at $(X/M_{BH},Y/M_{BH})=(10^{11},0)$ with impact parameters $b/M_{BH} \in [4\times10^2,4\times10^4]$. The BH is located at $(X,Y)=(0,0)$, indicated by the small solid dot. The size of the solid dot does not represent the physical size of the BH. Darker trajectories correspond to smaller impact parameters (i.e., photons passing closer to the BH), while brighter trajectories have larger impact parameters. The thicker curves highlight representative photon trajectories that lead to image multiplication: light emitted from the yellow star is received by the observer along two distinct photon paths with different impact parameters, and therefore appears at two different observation angles.}
    \label{fig. multiple sec image}
\end{figure*}

To illustrate the origin of multiple secondary images, we perform a backward ray-tracing procedure, computing photon trajectories from $D_{OL}/M_{BH}=10^{11}$ toward the BH and extending to infinity for a range of representative impact parameters. In the physical interpretation, these light rays are emitted from arbitrary sources toward the BH and then propagate to the observer's location. For this setup, we use the Hernquist EC DM with $M_{DM}/M_{BH}=100$ and $a_0/M_{BH}=10^4$. The resulting photon trajectories are shown in Fig.~\ref{fig. multiple sec image}.

One might notice the sharp turn of light rays occurring at $X/M_{BH} \sim 0$. This happens because the plot scale is extremely large and highly unequal, being of order $\sim 10^6 M_{BH}$ on the horizontal ($x$-)axis and $\sim 10^4 M_{BH}$ on the vertical ($y$-)axis. As a result, small changes along the $y$-axis appear significantly larger than those along the $x$-axis. The clear, smooth bending of light, as illustrated in Fig.~\ref{fig. lensing illus}, occurs much closer to the BH horizon at $(X,Y)=(0,0)$, at radii satisfying $\sqrt{X^2 + Y^2} = 2M_{BH}$. Therefore, at the scale of the plot, the light bending appears as a sharp turn at $X/M_{BH} \sim 0$. This choice of scale is made solely to illustrate more clearly the origin of image multiplication of a source, e.g., the yellow star.

It is evident that there exist specific ranges of smaller impact parameters (darker red curves) whose deflection is weaker than that of some photons with larger impact parameters (brighter orange–yellow curves). Consequently, photons belonging to these lower-impact-parameter ranges can intersect those with higher impact parameters at certain locations away from the BH (e.g., at the position of the yellow star in Fig.~\ref{fig. multiple sec image}). Since the impact parameter is directly related to the observation angle $\Theta^{(\pm)}$ (cf. Eq.~\eqref{eq. Theta pm}), the same source can be detected at (at least) two distinct angles. This leads to the formation of multiple (secondary) images. We did not investigate this behavior in further detail, as it lies beyond the scope of the present study.

\end{document}